\documentclass{article} %
\usepackage{colm2024_conference}

\usepackage{microtype}
\usepackage[hidelinks]{hyperref}
\usepackage{url}
\usepackage{booktabs}

\usepackage{amssymb}
\usepackage{amsmath}
\usepackage{tabularx}
\usepackage{multirow}
\usepackage{caption}
\usepackage{subcaption}
\usepackage[capitalize,noabbrev]{cleveref}
\usepackage{algorithm}
\usepackage{booktabs}       %
\usepackage{diagbox}
\usepackage{xspace}
\usepackage{listings}
\usepackage{wrapfig}

\usepackage[normalem]{ulem}

\usepackage{nicefrac}
\usepackage{graphicx}
\graphicspath{ {./graphs/} }
\usepackage{pgf}
\usepackage{xcolor}

\definecolor{codegreen}{rgb}{0,0.6,0}
\definecolor{codegray}{rgb}{0.5,0.5,0.5}
\definecolor{codepurple}{rgb}{0.58,0,0.82}
\definecolor{backcolour}{rgb}{0.95,0.95,0.92}

\lstdefinestyle{mystyle}{
    backgroundcolor=\color{backcolour},   
    commentstyle=\color{codegreen},
    keywordstyle=\color{magenta},
    numberstyle=\tiny\color{codegray},
    stringstyle=\color{codepurple},
    basicstyle=\ttfamily\footnotesize,
    breakatwhitespace=false,         
    breaklines=true,                 
    captionpos=b,                    
    keepspaces=true,                 
    numbers=left,                    
    numbersep=5pt,                  
    showspaces=false,                
    showstringspaces=false,
    showtabs=false,                  
    tabsize=2
}

\newcommand{\com}[1]{}
\newcommand{\resolved}[1]{}

\newcommand{\passk}[0]{pass@$k$\xspace}
\newcommand{\passflops}[0]{pass$_{\text{flops}}$@$f$\xspace}
\newcommand{\passtime}[0]{pass$_{\text{time}}$@$t$\xspace}
\newcommand{\rankk}[0]{rank-score@$k$\xspace}
\newcommand{\rankflops}[0]{rank-score$_{\text{flops}}$@$f$\xspace}
\newcommand{\ranktime}[0]{rank-score$_{\text{time}}$@$t$\xspace}

\usepackage{etoolbox}
\newtoggle{release}

\togglefalse{release}

\iftoggle{release}{
\newcommand{\tal}[1]{}
\newcommand{\michael}[1]{}
\newcommand{\roy}[1]{}
\newcommand{\adios}[1]{}
}
{
\newcommand{\tal}[1]{{\textcolor{cyan}{\textbf{Tal:} {#1}}}}
\newcommand{\michael}[1]{{\textcolor{orange}{\textbf{Michael:} {#1}}}}
\newcommand{\roy}[1]{{\textcolor{blue}{\textbf{Roy:} {#1}}}}
\newcommand{\adios}[1]{{\textcolor{magenta}{\textbf{YA:} {#1}}}}
}

\title{
The Larger the Better? Improved LLM Code-Generation via Budget Reallocation}

\author{Michael Hassid$^{1,2}$\thanks{~Equal contribution}~, Tal Remez$^{1 *}$, Jonas Gehring$^{1}$, Roy Schwartz$^{2}$, Yossi Adi$^{1,2}$\\
$^1$FAIR Team, Meta\\
$^2$The Hebrew University of Jerusalem\\
\texttt{\{michael.hassid\}@mail.huji.ac.il}\\}

\colmfinalcopy %
\begin{document}

\maketitle

\begin{abstract}
It is a common belief that \textit{large} language models (LLMs) are better than smaller-sized ones. However, larger models also require significantly more time and compute during inference. This begs the question: \emph{what happens when both models operate under the same budget?} (e.g., compute, run-time).
To address this question, we analyze code generation LLMs of various sizes and make comparisons such as running a $70$B model once vs.~generating five outputs from a $13$B model. We consider a standard unit-test setup, which can be used to select the correct output from the smaller model. Our findings reveal that the repeated use of smaller models can yield consistent improvements, with gains of up to $15$\% across five tasks. On the other hand, in scenarios where unit-tests are unavailable, a ranking-based selection of candidates from the smaller model falls short of the performance of a single output from larger ones. Our results highlight the potential of using smaller models instead of larger ones, and the importance of studying approaches for ranking LLM outputs.\footnote{\label{note_data} Data is avalible at \url{https://github.com/slp-rl/budget-realloc}}
\end{abstract}

\section{Introduction}
\begin{figure}[t]
     \centering
     \includegraphics[trim={0.65cm 5.7cm 3cm 0.05cm},clip,width=\textwidth]{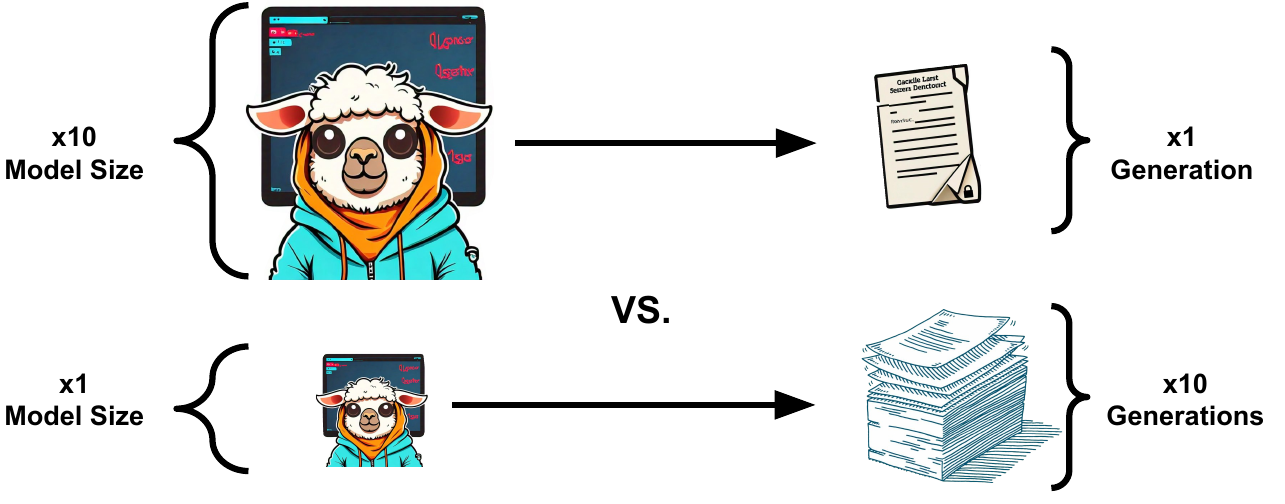}
    
     \caption{Different ways to improve LLM performance by increasing compute budget. Top: the standard approach of increasing model size, while generating a single output. Bottom: our approach---using a small model to generate multiple outputs, and select the best one.  
\label{fig:fig_1}}
\end{figure}

A common wisdom in deep learning, and language modeling in particular, is that investing more compute leads to improved performance \citep{kaplan2020scaling}. The standard way of implementing this principle is training larger models.
A simpler, yet often overlooked way to increase compute budget is to run a smaller model multiple times, and select the best output using some metric~\citep{humaneval}. In this work we systematically compare between these two approaches: we ask whether, given a fixed compute budget, it is best to run a large model once, or a smaller model multiple times~(\cref{fig:fig_1}). Our results show that, perhaps surprisingly, given the same compute budget, running 7B or 13B models can not only match the performance of a 70B model, but also substantially surpass it.

Addressing our research question requires a method for selecting the best LLM output from a given set of candidates. In this work we focus on execution based code-generation tasks, which assume the availability of unit-tests~\citep{humaneval,mbpp,apps}. We consider the widely-used \passk metric~\citep{spoc}, which evaluates a model's performance on code generation problems by generating $k$ outputs and assigning a point if any of them passes all tests. To adapt this metric for our purposes, we take models of different sizes, and for each generate as many outputs as possible given a fixed compute budget, e.g., floating point operations (FLOPs) or wall-time.

We apply this setup to evaluate the Code Llama \citep{roziere2023code} model family ($7$B, $13$B, $34$B, and $70$B) across five tasks: HumanEval \citep{humaneval}, MBPP \citep{mbpp}, and the three splits of APPS \citep{apps}. For the HumanEval and MBPP benchmarks, we additionally use the recent Llama-3 \citep{llama3modelcard} model family ($8$B and $70$B). Surprisingly, we find that for the two popular tasks, HumanEval and MBPP, the smaller models ($7$B, $8$B and $13$B) outperform the larger ones ($34$B and $70$B) by a margin of up to $15$\%. Importantly, this is observed using both budget types (FLOPs and wall-time) and across all computation budgets. When considering the challenging APPS benchmark, we find that the $13$B model performs best across almost all budgets, with a consistent margin of $5$\% when considering the hardest split---competition. 

We then proceed to examine the scenario where unit-tests are \textit{unavailable}, such as in an IDE code-completion setup. In such cases, an efficient policy is required to select a single solution from all generated ones. We consider a simple LLM-based policy, which ranks solutions based on the negative log likelihood of the LLM. We also augment this policy with a variant of a recent ranking approach---LEVER \citep{ni2023lever}. We experiment with the $7$B model, and rank its outputs using each of the models. 
Our results show that, as expected, ranking-based selection improves with the increase in compute budget, and with the size of the ranking LLM. Nonetheless, this procedure still falls short of the performance achieved by running the larger model independently with the same budget.

Our results highlight the potential of using smaller models instead of larger ones, a practice that has many benefits. First, small models are far computationally cheaper to pre-train.\footnote{E.g., Llama-$2$ $7$B was $\approx$$10X$ faster to pre-train compared to the $70$B variant~\citep{touvron2023llama}.} Further, at inference time, they are considerably more hardware-friendly: a $13$B model can be accommodated on a single A$100$ GPU, a feat unachievable for a $70$B model~\citep{dettmers2022gpt3}. 
Finally, as we have shown, when controlling for the compute budget, smaller models may actually outperform larger ones. 

Our findings also emphasize the importance of developing effective ranking approaches for LLM outputs. This is especially important in cases where no unit-tests or other verification methods are available~\citep{zou2021pre, uesato2022solving, sun2023chatgpt}. To support such research direction, we release $2,000$ Code~Llama~$7$B outputs for each example in HumanEval and MBPP---a total of more than $1$M outputs.\footnotemark[1]

\label{intro}
\section{Evaluation under Compute Restrictions}
\label{sec:eval_metrics}

To study our main research question---what is the optimal way of using a given LLM compute budget---we consider a code-generation setup with unit-tests~\citep{humaneval,mbpp,apps}. 
Below we discuss our methodology for code generation evaluation under computational restrictions. We begin by describing \passk \citep{spoc}, the current main approach for evaluating code generation tasks~(\cref{subsec:pass_at_k}). We then transition to describe our variant of code generation metrics under computational restrictions~(\cref{subsec:pass_at_comp}). 

\subsection{Standard Code Generation Evaluation}
\label{subsec:pass_at_k}
To evaluate LLM code-generation abilities, a common setup assumes a set of coding questions, each with a set of unit-tests. The LLM is fed with each question, and a fixed number of output generations (labelled $k$) are sampled. The evaluation  protocol considers each question for which at least one output passes all unit-tests as correct. To estimate the performance of a model that generates $k$ outputs, it is common to generate a larger number of outputs $n$ (\(>k\)) and compute:
\begin{align}
\text{\passk} &:= \mathop{\mathbb{E}}_{\text{Problems}} \left[ 1 - \frac{{\binom{n-c}{k}}} {\binom{n}{k}} \right],
\label{eq:pass_at_k}
\end{align}

where \(c \leq n\) is the number of examples that pass the unit-tests. The above mentioned metric results in an unbiased estimator as was shown by~\cite{humaneval}.

\subsection{Comparing LLMs of Different Sizes under a Fixed Budget}
\label{subsec:pass_at_comp}
Our goal is to compare between LLMs of different sizes under a fixed compute budget. To do so, we allow smaller models, which consume fewer resources, to generate more outputs. This results in models of different sizes requiring roughly the same amount of compute.

We consider two types of compute budgets: the number of FLOPs and wall-time. For each type, a specific resource limit is set (e.g., $10$k Tera-FLOPs or $8$ seconds), and the model generates examples up to the point where the compute limit is reached.
That is:
\begin{align}
\text{\passflops} &:= \text{\passk} \quad \text{where: } k=\max_{\text{flops($k'$)} \leq f} k',  
\label{eq:pass_at_flops} \\
\text{\passtime} &:= \text{\passk} \quad \text{where: } k=\max_{\text{time($k'$)} \leq t} k',  \label{eq:pass_at_time}
\end{align}

where flops($k$) and time($k$)  are functions that return the FLOPs/wall-time usage of a given model that generates $k$ outputs.
Notably, the FLOPs restriction is a more theoretical computational restriction, as it assumes perfect utilization of the hardware. On the other hand, the wall-time restriction is more realistic, but is hardware specific, and thus not directly comparable across different machines.

\section{Experimental Setup}
\label{sec:exp_setup}
In this section we describe our experimental setup, focusing on the code benchmarks used~(\cref{subsec:datasets}), our metrics~(\cref{subsec:metrics_setup}), and our experiments~(\cref{subsec:setup}).

\subsection{Benchmarks}
\label{subsec:datasets}
We experiment with three python code benchmarks: HumanEval~\citep{humaneval}, MBPP~\citep{mbpp} and APPS~\citep{apps}.
The HumanEval benchmark consists of 164 function declarations alongside their documentation. The Code-LLM's task is to complete the function according to the provided documentation.
MBPP consists of $500$ test examples, each one is an instruction for a code function. Here, the Code-LLM is required to generate the full function. 
Lastly, the test subset of APPS is composed of $5$k programming problems at
various levels of difficulty: introductory~($1$k), interview~($3$k) and competition~($1$k). In the APPS tasks, the Code-LLM is required to generate the complete python file, which includes import declarations, class definitions, and so on.

\subsection{Metrics}
\label{subsec:metrics_setup}
Computing the \passflops and \passtime metrics requires an estimation of the flops($k$) and time($k$) functions from \cref{eq:pass_at_flops,eq:pass_at_time}. To estimate FLOPs usage, we use the calflops library \citep{calflops}, with input sequence length of $128$. 
We measure wall-time while assuming optimal throughput utilization of the hardware. Specifically, we use a node of $8$ A$100$ GPUs, optimize the batch size per model and measure the time it takes each model to generate a subset of $\approx$1k examples from our datasets. We report the Code Llama results in \cref{tab:factors}, for readability we also report the normalized factor with respect to the $7$B model.\footnote{Llama-3 $8$B/$70$B presents  similar usage to Code Llama $7$B/$70$B, with a difference of up to 7\%.}

\begin{table}
\caption{\label{tab:factors} Code Llama FLOPS and wall-time usage per model size, along with normalized values with respect to the $7$B model. }
\begin{center}
\begin{tabularx}{\textwidth}{l|cc|cc}
\toprule
Model Size & FLOPs (Teras) & FLOPs (norm.) & wall-time (seconds) & wall-time (norm.) \\
\midrule
\phantom{00}7B  & 1.69  & 1.00  & 395   & 1.00 \\
\phantom{0}13B & 3.29  & 1.95  & 667   & 1.69 \\
\phantom{0}34B & 8.58  & 5.08  & 2,994\phantom{0} & 7.58 \\
\phantom{0}70B & 17.60\phantom{0} & 10.41\phantom{0} & 5,605\phantom{0} & 14.19\phantom{0} \\
\bottomrule
\end{tabularx}
\end{center}
\end{table}

\subsection{Experiments}
\label{subsec:setup}
We experiment with the Code Llama family~\citep{roziere2023code}, a finetuned version of Llama~\citep{touvron2023llama}. Code Llama comes in various sizes, which we use for our experiments: $7$B, $13$B, $34$B and $70$B. For the smaller benchmarks, HumanEval and MBPP, we also consider the Llama-3 family ($8$B and $70$B).

We follow \citet{roziere2023code}, and use a zero-shot setting for HumanEval, a $3$-shot prompting strategy for MBPP and $2$-shot prompts for APPS, and limit the generation length to $512$/$256$/$256$ tokens for HumanEval/MBPP/APPS.
For the sampling process, we use nucleus sampling~\citep{holtzman2019curious} with top-$p=0.95$ and a temperature of $0.8/0.8/0.6$ for HumanEval/MBPP/APPS, with all models sizes~\citep{roziere2023code}. Finally, we also report, the pass@1 results using a greedy decoding method for all models.

To compare models in varying sizes, we select the maximal number of generations for each model with respect to the values in \cref{tab:factors}. Specifically, for the smaller benchmarks, HumanEval and MBPP, we generate $n = 2,000/1,000/400/200$ answers for the $7$-$8$B/$13$B/$34$B/$70$B models, respectively. For the larger benchmarks, the three splits of APPS, we use $n = 1,000/500/200/100$. To get a robust estimation of these measures, we follow \cite{humaneval} and \cite{roziere2023code}, and report for all benchmarks a maximal value of $k=\frac{n}{2}$ for the \passk metric, while using all available unit-tests.

\section{Small Models Outperform Large Ones under a Fixed Compute Budget}
\label{sec:results}

\begin{figure}[t]
     \centering
     \begin{subfigure}[b]{0.32\textwidth}
         \centering
         \includegraphics[trim={0.7cm 0.3cm 0.7cm 0.7cm},clip,width=\textwidth]{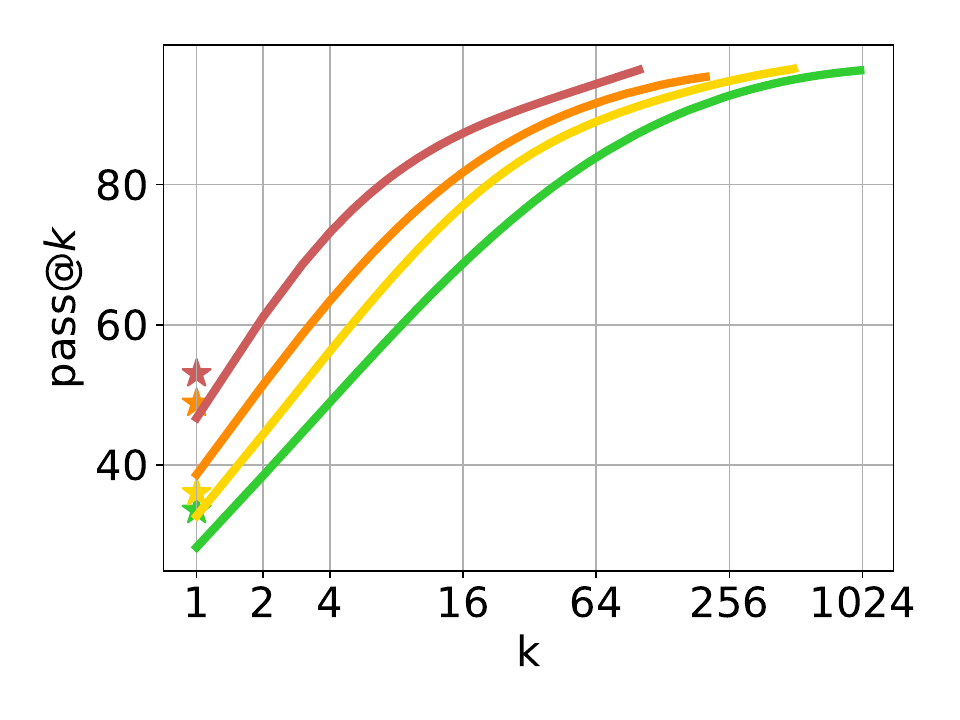} 
         \caption{\label{fig:human_eval_k}}
     \end{subfigure}
     \hfill
     \begin{subfigure}[b]{0.32\textwidth}
         \centering
         \includegraphics[trim={0.7cm 0.3cm 0.7cm 0.7cm},clip,width=\textwidth]{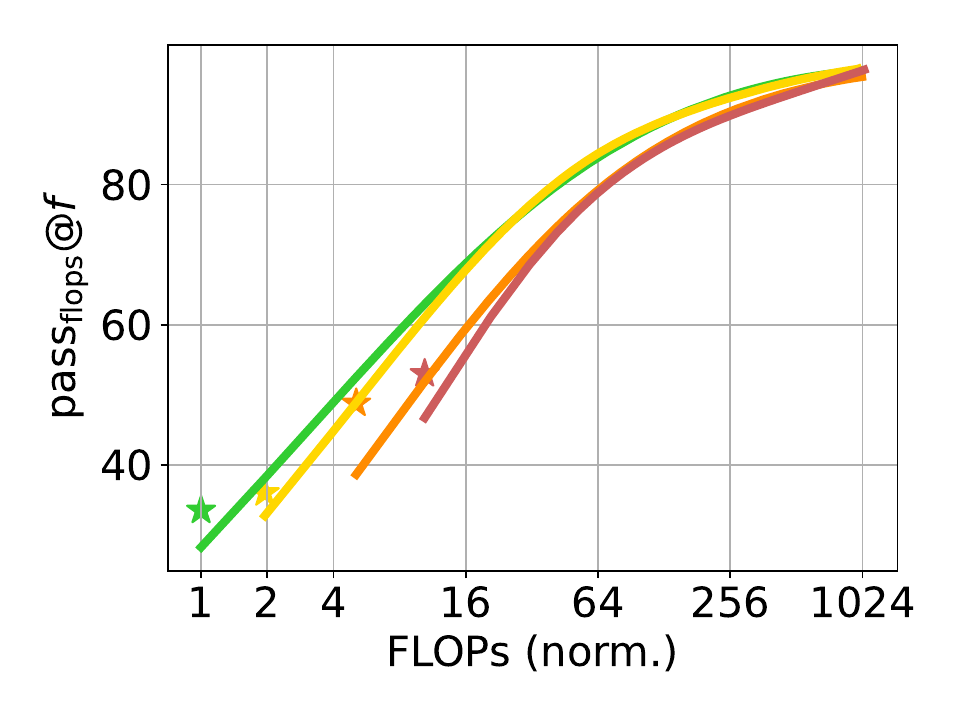} 
         \caption{\label{fig:human_eval_flops}}
     \end{subfigure}
     \hfill
     \begin{subfigure}[b]{0.32\textwidth}
         \centering
         \includegraphics[trim={0.7cm 0.3cm 0.7cm 0.7cm},clip,width=\textwidth]{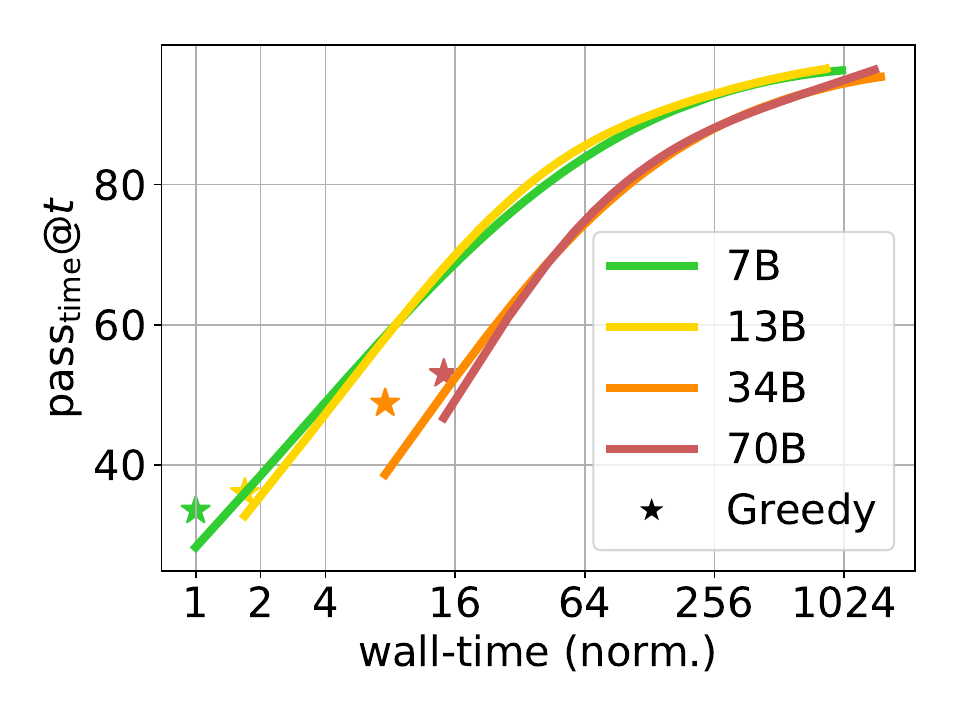} 
         \caption{\label{fig:human_eval_time}}
     \end{subfigure}
     \caption{Code Llama performance (Y axis) as a function of compute (X axis, in exponential scale) for the HumanEval benchmark. Larger models perform better in general (\cref{fig:human_eval_k}), but under a fixed compute budget (\cref{fig:human_eval_flops,fig:human_eval_time}),  smaller models ($7$B and $13$B)  substantially outperform larger ones ($34$B and $70$B). Greedy decoding is marked by a star.
\label{fig:human_eval}}
\end{figure}

\begin{figure}[t]
     \centering
     \begin{subfigure}[b]{0.32\textwidth}
         \centering
         \includegraphics[trim={0.7cm 0.3cm 0.7cm 0.7cm},clip,width=\textwidth]{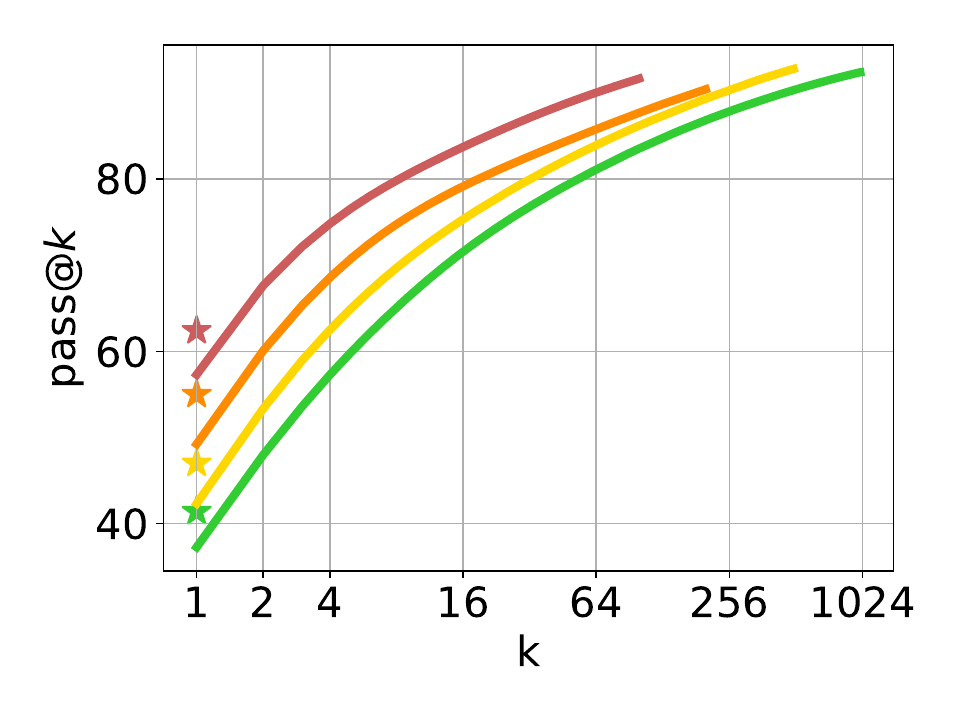}
         \caption{\label{fig:mbpp_k}}
     \end{subfigure}
     \hfill
     \begin{subfigure}[b]{0.32\textwidth}
         \centering
         \includegraphics[trim={0.7cm 0.3cm 0.7cm 0.7cm},clip,width=\textwidth]{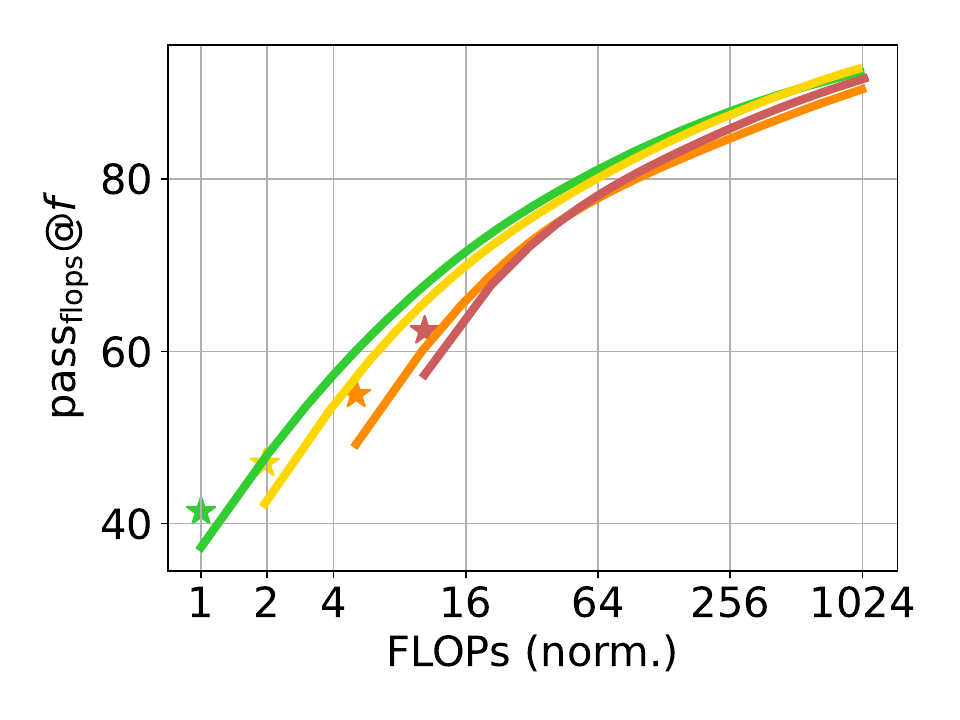}
         \caption{\label{fig:mbpp_flops}}
     \end{subfigure}
     \hfill
     \begin{subfigure}[b]{0.32\textwidth}
         \centering
         \includegraphics[trim={0.7cm 0.3cm 0.7cm 0.7cm},clip,width=\textwidth]{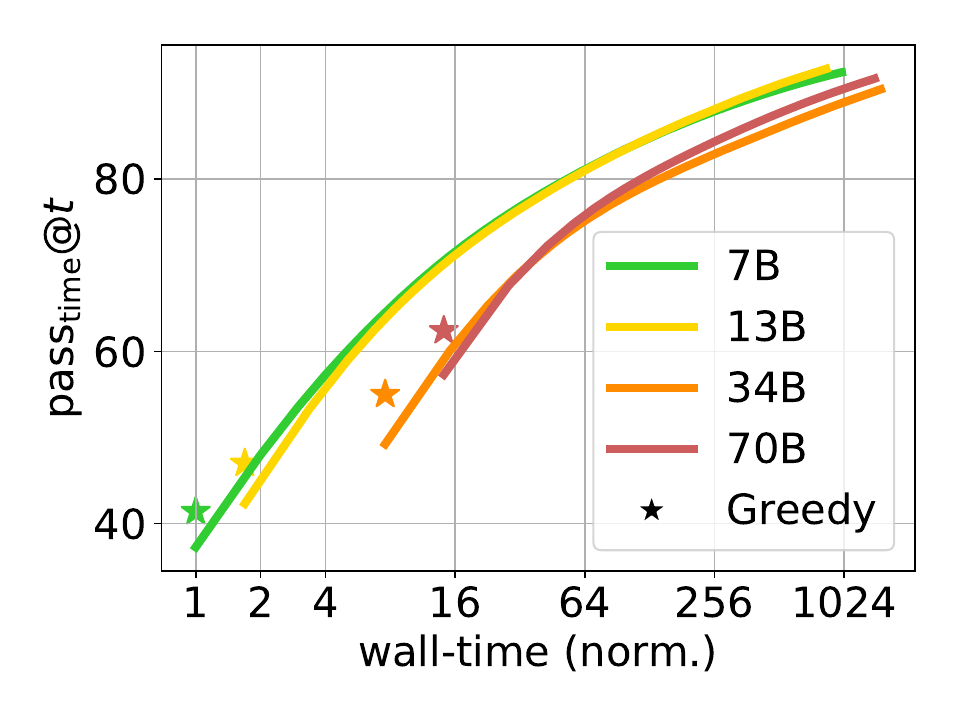}
         \caption{\label{fig:mbpp_time}}
     \end{subfigure}
     \caption{Code Llama performance vs.~compute for the MBPP benchmark. As in HumanEval~(\cref{fig:human_eval}), larger models perform better as a function of $k$ (\cref{fig:mbpp_k}), but worse under a fixed compute budget (\cref{fig:mbpp_flops,fig:mbpp_time}). %
\label{fig:mbpp}}
\end{figure}

Results for \textbf{HumanEval} and \textbf{MBPP} using the Code Llama models are presented in \cref{fig:human_eval,fig:mbpp}, respectively.\footnote{\cref{tab:full_human_eval,tab:full_mbpp} in \cref{app:full_results} presents detailed results.} The  corresponding results for the Llama-3 models can be found in \cref{fig:llama_3_human_eval,fig:llama_3_mbpp} (\cref{app:llama_3}).
We first note that, as expected, the \passk metric improves both with model scale, and with the number of generations $k$~(sub-figure (a) in all figures).
However, perhaps surprisingly, when considering the \passflops and \passtime metrics~
(sub-figures (b) and (c)), we see a different trend---given a fixed compute budget, smaller models yield better results than larger ones. Specifically, the $7$B/$8$B/$13$B models outperform the larger models across all compute budgets. Particularly, in the small budget regime (up to $32$ normalized FLOPs units and $64$ wall-time units) the performance gap reaches $5$---$15\%$.

Another way of looking at our results is by observing that smaller models match the performance of larger ones using substantially lower budgets. For instance, in HumanEval, the Code Llama $7$B and $13$B models achieve a score of $60\%$ using one quarter of the time it takes the larger models to reach that score. This efficiency gap further increases with the Llama-3 models~(\cref{fig:llama_3_human_eval_time}). Finally, we compare small models to greedy decoding with larger models, which generally performs better than sampling. We observe that even in this setup, using the smaller models several times is equivalent or preferable in all cases.

\begin{figure}[t!]
     \centering
     \begin{subfigure}[b]{0.32\textwidth}
         \centering
         \includegraphics[trim={0.7cm 0.3cm 0.7cm 0.7cm},clip,width=\textwidth]{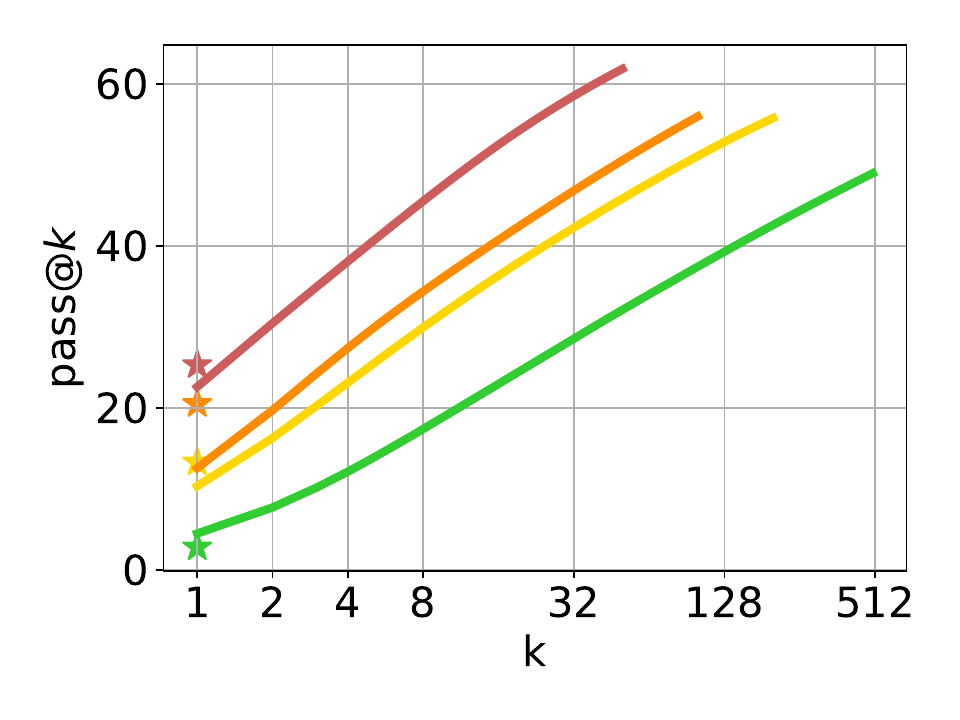}
         \caption{\label{fig:apps_intro_k}}
     \end{subfigure}
     \hfill
     \begin{subfigure}[b]{0.32\textwidth}
         \centering
         \includegraphics[trim={0.7cm 0.3cm 0.7cm 0.7cm},clip,width=\textwidth]{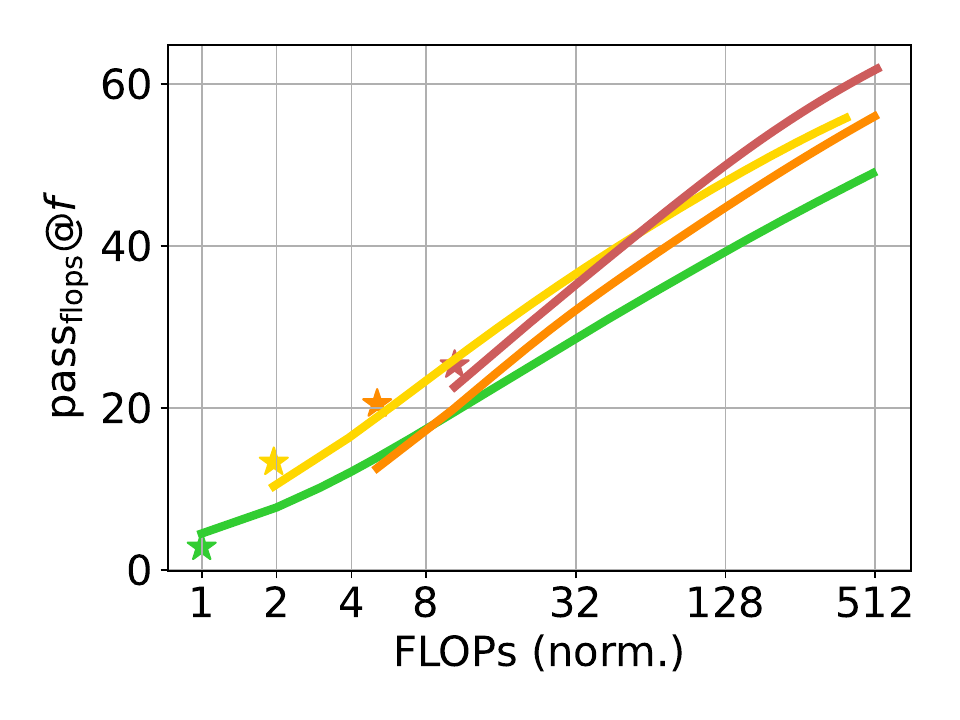}
         \caption{\label{fig:apps_intro_flops}}
     \end{subfigure}
     \hfill
     \begin{subfigure}[b]{0.32\textwidth}
         \centering
         \includegraphics[trim={0.7cm 0.3cm 0.7cm 0.7cm},clip,width=\textwidth]{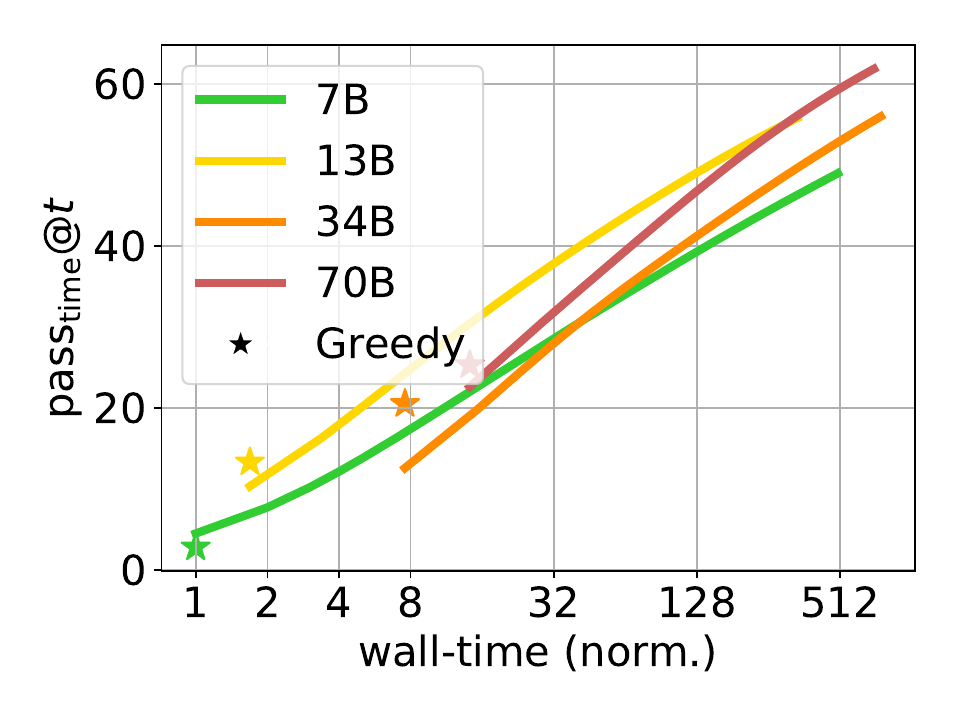}
         \caption{\label{fig:apps_intro_time}}
     \end{subfigure}
     \caption{Code Llama performance vs.~compute for the APPS benchmark, \textbf{introductory} split. 
     The $13$B model is superior to the $34$B model and comparable to the $70$B model under fixed budget.
     In contrast, the $7$B model underperforms the larger models.\label{fig:apps_intro}}
\end{figure}

\begin{figure}[t!]
     \centering
     \begin{subfigure}[b]{0.32\textwidth}
         \centering
         \includegraphics[trim={0.7cm 0.3cm 0.7cm 0.7cm},clip,width=\textwidth]{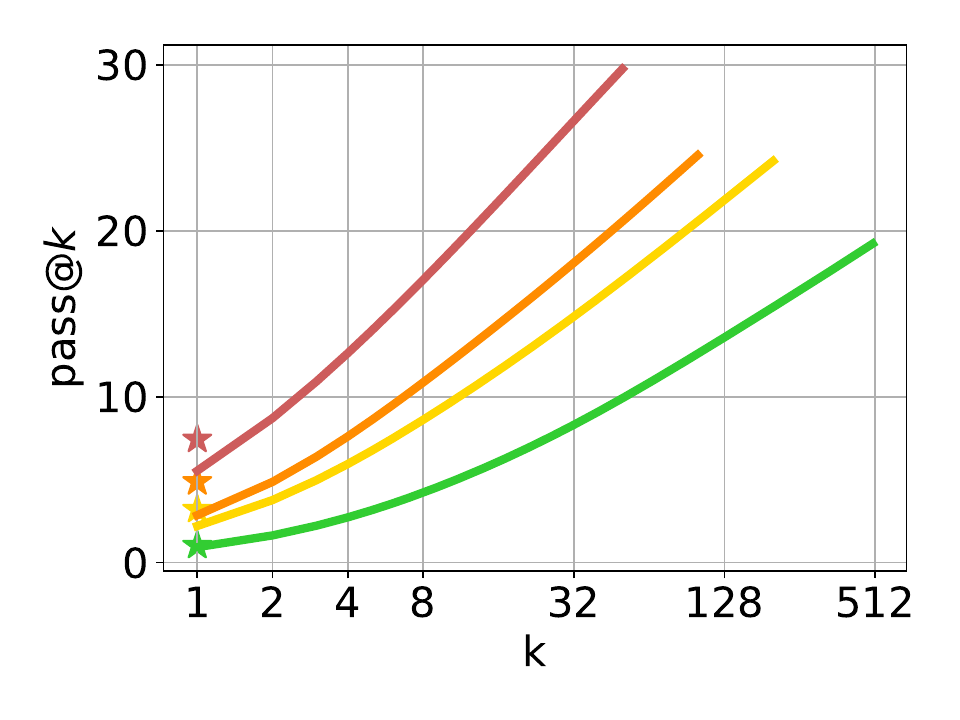}
         \caption{\label{fig:apps_interview_k}}
     \end{subfigure}
     \hfill
     \begin{subfigure}[b]{0.32\textwidth}
         \centering
         \includegraphics[trim={0.7cm 0.3cm 0.7cm 0.7cm},clip,width=\textwidth]{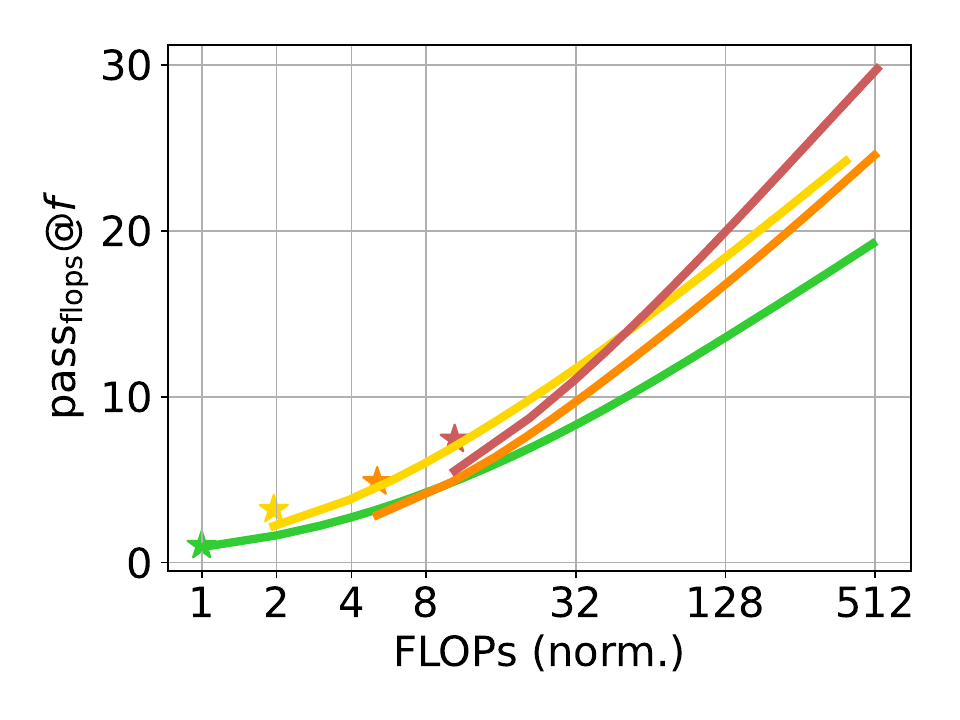}
         \caption{\label{fig:apps_interview_flops}}
     \end{subfigure}
     \hfill
     \begin{subfigure}[b]{0.32\textwidth}
         \centering
         \includegraphics[trim={0.7cm 0.3cm 0.7cm 0.7cm},clip,width=\textwidth]{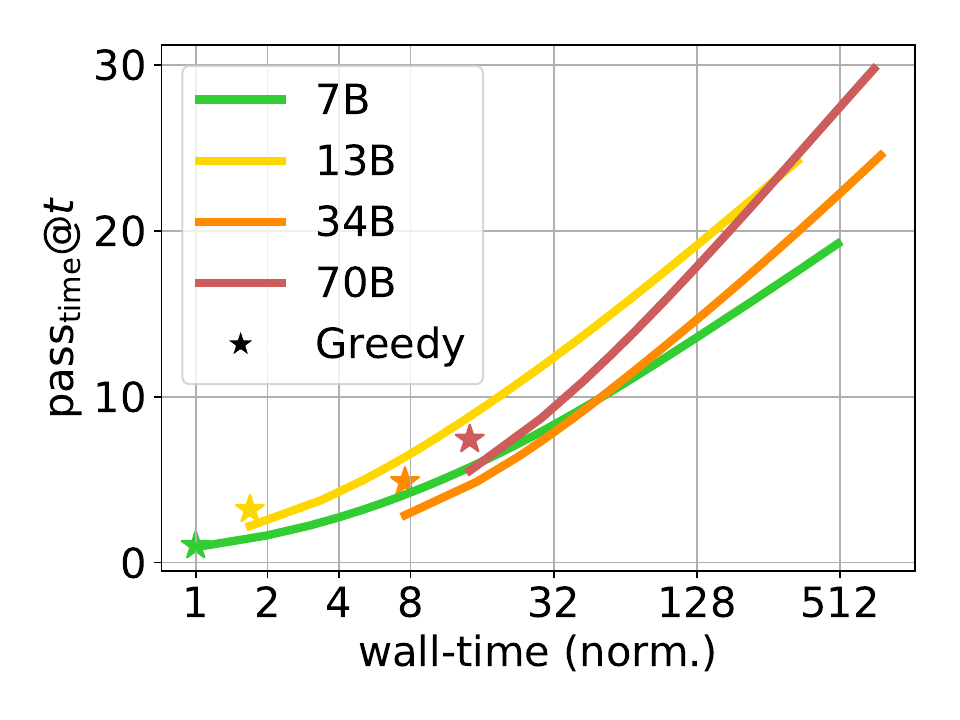}
         \caption{\label{fig:apps_interview_time}}
     \end{subfigure}
     \caption{Code Llama performance vs.~compute for the APPS benchmark, \textbf{interview} split. Similarly to the introductory split (\cref{fig:apps_intro}), the $13$B model is superior to the $34$B model and comparable to the $70$B model under fixed wall-time, while the $7$B model is inferior to the larger models.\label{fig:apps_interview}}
\end{figure}

\begin{figure}[t!]
     \centering
     \begin{subfigure}[b]{0.32\textwidth}
         \centering
         \includegraphics[trim={0.7cm 0.3cm 0.7cm 0.7cm},clip,width=\textwidth]{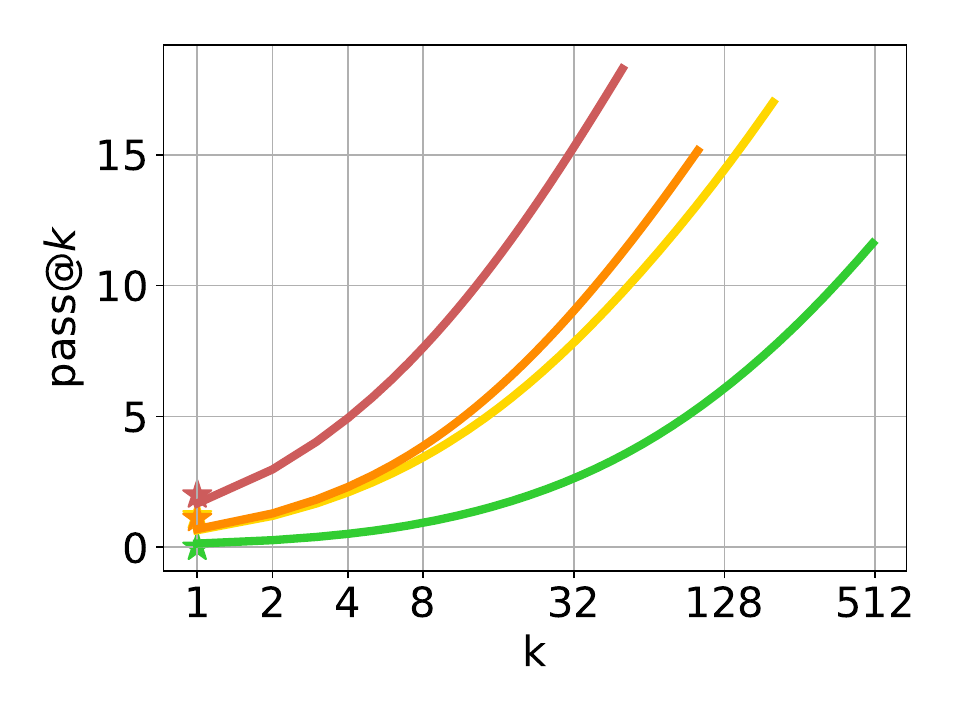}
         \caption{\label{fig:apps_comp_k}}
     \end{subfigure}
     \hfill
     \begin{subfigure}[b]{0.32\textwidth}
         \centering
         \includegraphics[trim={0.7cm 0.3cm 0.7cm 0.7cm},clip,width=\textwidth]{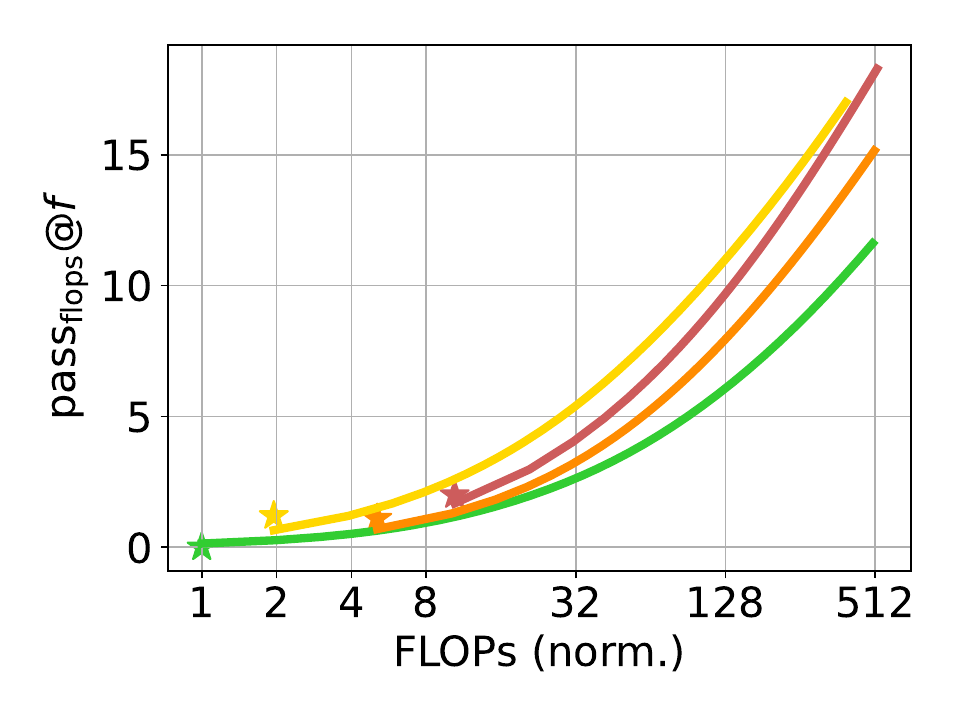}
         \caption{\label{fig:apps_comp_flops}}
     \end{subfigure}
     \hfill
     \begin{subfigure}[b]{0.32\textwidth}
         \centering
         \includegraphics[trim={0.7cm 0.3cm 0.7cm 0.7cm},clip,width=\textwidth]{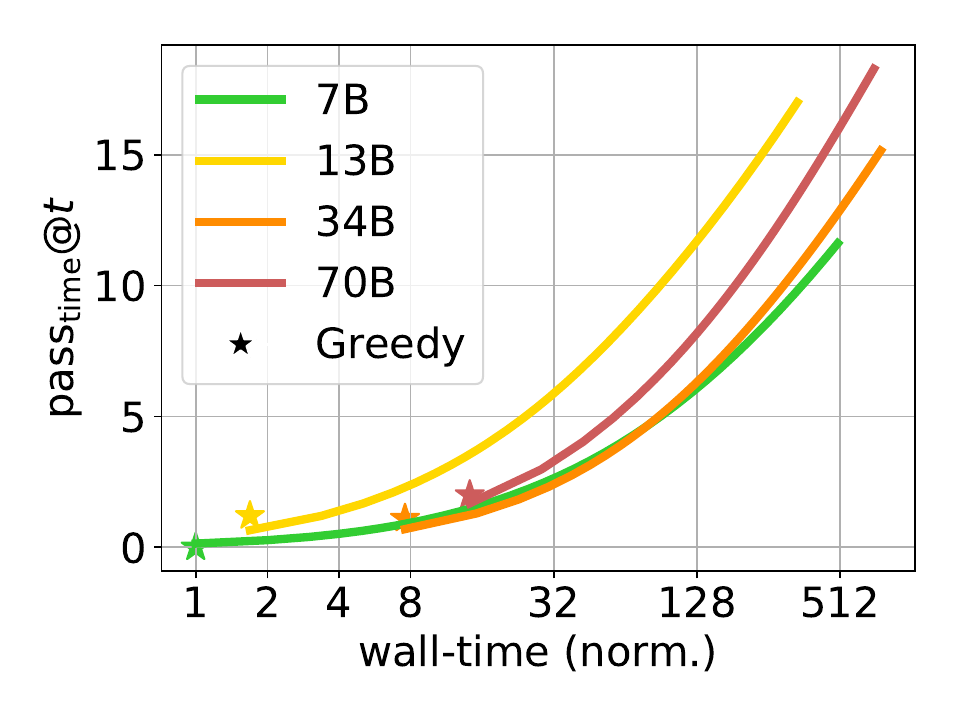}
         \caption{\label{fig:apps_comp_time}}
     \end{subfigure}
     \caption{Code Llama performance vs.~compute for the APPS benchmark, \textbf{competition} split (the most challenging one). The $13$B model is superior to both $34$B and $70$B models under fixed wall-time, and comparable to the $70$B under fixed number of FLOPs.\label{fig:apps_comp}}
\end{figure}

We next turn to discuss the Code Llama results over the three splits of the \textbf{APPS} benchmark~(\cref{fig:apps_intro,fig:apps_interview,fig:apps_comp}).\footnote{\cref{tab:full_apps} in \cref{app:full_results} presents detailed results.} We first consider the $13$B model, and observe the same trends as in HumanEval and MBPP: this model achieves the best performance in almost all fixed compute budgets. Specifically for the competition split~(\cref{fig:apps_comp_flops,fig:apps_comp_time}), the most challenging APPS split, the $13$B model outperforms all other models in all compute budgets, with a consistent margin of $\approx$$5\%$ from the $70$B model when considering the wall-time budget.
We further observe that the $13$B model achieves similar or better performance as the greedy approach of all models in all three splits. Finally, when fixing the performance, the $13$B model is $2$--$4$ times more efficient than the $70$B model (both for FLOPs and wall-time).

We next observe that the $7$B model is also competitive with larger models in small budget regimes (up to $8$ normalized FLOPs units and $16$ wall-time units). Nonetheless, it slightly underperforms the other models on larger budgets.
This can be attributed to the $7$B model's inability to generate a sufficient number of correct answers for the task, and may suggest that there is a minimum size requirement for a certain level of task difficulty.

Our results indicate that small models can match or even outperform large ones under a fixed compute budget, assuming the availability of unit-tests.
An intriguing aspect of our research question is what happens when unit-tests are \textit{unavailable}, and a single selection among several generations must be made. We delve into this topic in the following section.

\section{Evaluating Code Generation without Unit-tests}
\label{sec:ranking}

We examine the scenario where unit-tests are not available (e.g., IDE code-completion setup). In this case, an efficient selection policy strategy may be used to select one answer from the model's generations. In the previous cases (\cref{sec:eval_metrics}), unit-tests served as this policy. 
Here we investigate using ranking as a selection policy. In \cref{subsec:eval_rankers} we show how to estimate the performance of a model given such a strategy, and in \cref{subsec:results_rankers} we analyze the performance of larger models as rankers for a small model.

\subsection{Evaluating Rankers}
\label{subsec:eval_rankers}
We assume a model that generates $k$ outputs, and a policy that ranks them. To estimate the performance of such setup, we count the number of groups containing $k$ generations where the highest-ranked generation within them is a correct one. That is:
\begin{align}
\text{\rankk} &:= \mathop{\mathbb{E}}_{\text{Problems}} \left[  \frac{1} {\binom{n}{k}} \cdot \left( \sum_{i=1}^{n-k+1}{\binom{n-i}{k-1} \cdot \text{pass}_i } \right) \right],
\label{eq:rank_at_k}
\end{align}

where $n(>$$k)$ is the number of answers generated for the estimation, and \([\text{pass}_1, \text{pass}_2, \dots, \text{pass}_n] \in \{0, 1\}^n\) are the pass scores sorted according to the ranking policy. That is, pass$_i$ is $1$ if the example ranked $i$ according to the policy is correct, and $0$ otherwise. See~\cref{fig:impl} for a python implementation of \rankk.

\begin{figure}[t!]
\centering
\begin{lstlisting}[language=Python,breaklines=true]
def rank_score_at_k(n, k, pass_sorted):
    """
    :param n: total number of samples
    :param k: k in rank-score@k
    :param pass_sorted: a binary list of pass scores. The list is sorted by the ranks assigned to examples by a ranker.
    """
    numerator_sum = 0
    for i in range(1, n-k+2):
        numerator_sum += math.comb(n-i, k-1) * scores_and_pass[i-1]
    score = (numerator_sum / math.comb(n, k)) * 100
    return score
\end{lstlisting}
\caption{A Python implementation of \rankk as presented in \cref{eq:rank_at_k}.}
\label{fig:impl}
\end{figure}

Similarly to \cref{eq:pass_at_flops,eq:pass_at_time}, we also define:
\begin{align}
\text{\rankflops} &:= \text{\rankk} \quad \text{where: } k=\max_{\text{flops($k'$)} \leq f} k',
\label{eq:rank_at_flops} \\
\text{\ranktime} &:= \text{\rankk} \quad \text{where: } k=\max_{\text{time($k'$)} \leq t} k',  \label{eq:rank_at_time}
\end{align}
where flops($k$) and time($k$) are the same functions as in \cref{subsec:pass_at_comp}.
Next, we evaluate the performance of large models as rankers using the above metrics.

\subsection{Large Language Models as Rankers}
\label{subsec:results_rankers}

We examine the usage of LLMs as rankers. To produce a ranking order over a set of generations, we use the averaged Negative Log Likelihood (NLL) the LLM assigns to each generation (excluding the prompt), and rank the generations according to that score. %
It should be noted that extracting the NLL of a model over a given generation can be done in a parallel manner (i.e., non-autoregressively), which is substantially faster than traditional token-by-token generation. The score given by a model to a generation $G=(w_1, \dots, w_l)$ given a prompt $P$ is:
\begin{align}
\text{score}_{model} = \text{NLL}_{model}(G | P) = -\frac{1}{l}\sum_{i=1}^{l} \log{\big(p_{model}(w_i | w_{i-1}, \dots, w_1, P)\big)}.
\label{eq:model_score} 
\end{align}

To study the performance of LLMs as rankers we use the HumanEval and MBPP benchmarks.
We use $2,000$ generations produced by Code Llama $7$B as described in \cref{subsec:setup}. As rankers we use all four Code Llama model sizes. We discard any generation that fails to complete, i.e. reached the maximal number of generated tokens without producing an end-of-sequence token. We also report the performance of running each model independently with one generation budget (both greedy and sampling).

\begin{figure}[t!]
     \centering
     \begin{subfigure}[b]{0.46\textwidth}
         \centering
         \includegraphics[trim={0.7cm 0.9cm 0.7cm 0.7cm},clip,width=0.91\textwidth]{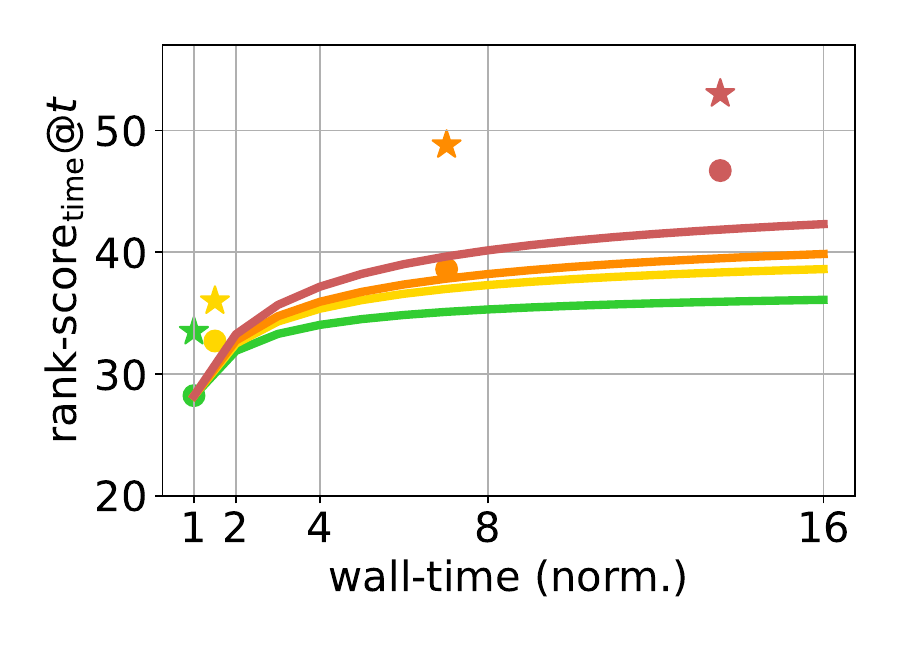}
     \end{subfigure}
     \hfill
     \begin{subfigure}[b]{0.46\textwidth}
         \centering
         \includegraphics[trim={0.7cm 0.9cm 0.7cm 0.7cm},clip,width=0.91\textwidth]{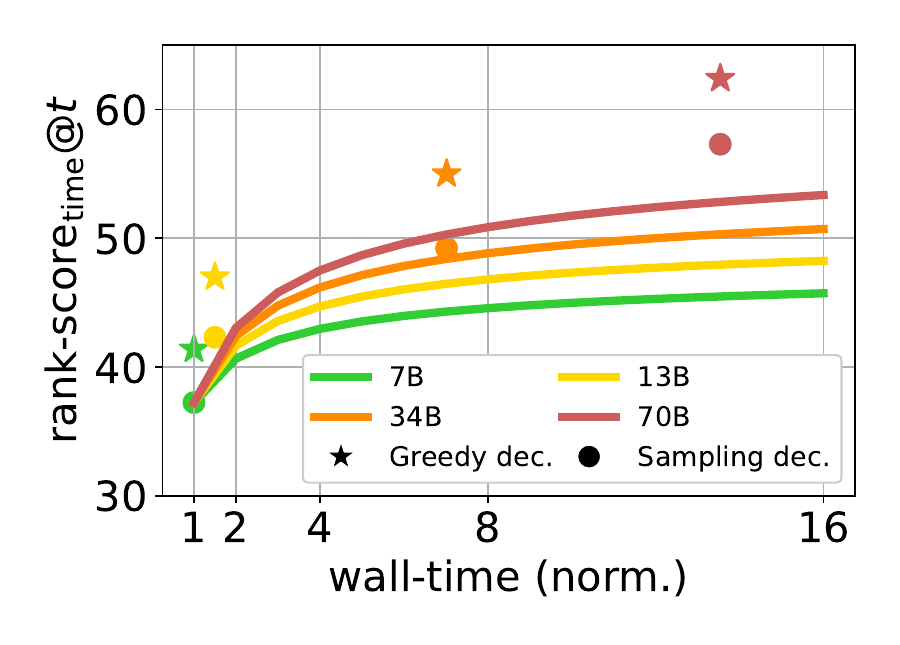}
     \end{subfigure}
     \caption{\text{\ranktime} as a function of wall-time for HumanEval (left) and MBPP (right), using different rankers (different lines). Greedy sampling is marked as a star, and top-p sampling as a circle. While ranking results improve with the size of the ranker and with compute budget, they still fall short of greedy decoding with larger models.\label{fig:rank}}
\end{figure}

Our results are presented in \cref{fig:rank}. As can be seen, using LLMs as rankers over generations obtained from smaller models improves performance. 
Interestingly, we observe that using a $7$B model as a ranker for itself can enhance its generation even further than the greedy approach, albeit with the cost generating several outputs. 
We also find that using larger models as rankers results in better perfomance.
When considering a fixed compute budget, we find that it is sometimes comparable to use LLMs as rankers instead of sampling from them, as can be seen with the $13$B and $34$B models. However, this is not the case for the greedy approach which consistently outperforms ranking multiple generations from a smaller model given a fixed compute budget. 

\begin{wrapfigure}{R}{0.37\textwidth}
  \centering
  \includegraphics[trim={0.7cm 0.9cm 0.7cm 0.7cm}, clip, width=0.33\textwidth]{graphs/rank_human_eval.pdf}
  \caption{\text{\ranktime} as a function of wall-time for MBPP, using the LEVER verfier with different NLL rankers. Results are similar to \cref{fig:rank}. \label{fig:lever}}
\end{wrapfigure}

To further check the use of external verifiers, we integrate the LEVER verifier model~\citep{ni2023lever} with the Code Llama models. The LEVER approach aims to enhance code generation by learning to verify generated programs. The full LEVER pipeline involves using the NLL produced by the code generation model, error pruning based on execution, and a verifier trained on code generations with execution results. However, since we assume that no tests are available in our setting, execution pruning and execution results cannot be used. LEVER released a trained verifier over the MBPP benchmark, which we use along with the NLL scores of each model. As shown in \cref{fig:lever}, the LEVER verifier does not improve the results in the test-less setting, which is expected given that one of the main components of the approach relies on execution over unit-tests.

In summary, there remains a gap to bridge between using LLMs as rankers for smaller models and using them as generators. To further promote this line of research, we release the $2,000$ generations per example produced by the $7$B model for both HumanEval and MBPP (a total of $1,328,000$ generations).

\section{Related Work}
\label{sec:related_work}

\subsection{Model Scaling}
Model scaling was found to be one of the key elements in the success of LLMs~\citep{dehghani2023scaling, gu2023scaling, hassid2024textually, rae2021scaling, chowdhery2023palm, touvron2023llama}, with \citet{wei2022emergent} demonstrating how specific abilities emerge mainly after reaching a specific scale. The way language models behave when they are scaled up and their ability to adjust have been a significant factor in the creation of LLMs~\citep{hernandez2021scaling}. \cite{kaplan2020scaling} investigated the optimal model size to train for a given compute budget, while \cite{hoffmann2022training} demonstrated how scaling both model and dataset sizes improves performance across various tasks. \cite{clark2022unified} analyzed the scaling properties of mixture-of-experts models, showing that scaling with the number of experts diminishes as model size increases. Recently, \cite{gadre2024language} provided a scaling law analysis considering downstream tasks rather than next-token prediction loss. They related the perplexity of a language model to its downstream task performance via a power law and used it to predict the top-1 error averaged over the evaluated downstream tasks. Our work differs from all of the above, as we do not claim to provide new scaling laws but rather suggest that when fixing the budget, smaller models can provide comparable or superior results to larger ones. 

Recent studies by \citet{shi2024need} and \citet{mei2024bigger} have demonstrated that under constrained compute budgets, smaller vision models can surpass their larger counterparts. Specifically, \citet{shi2024need} found advantages in using multiple image scales, whereas \citet{mei2024bigger} observed that smaller diffusion models perform better than larger ones when the compute budget is fixed. Our approach, which generates multiple text outputs from a small model, aligns with these findings.

\subsection{Verifiers and Rankers}
LLM verifiers and rankers is a growing trend, which leverages LLMs to verify and rank generations obtained from weaker and smaller models~\citep{cobbe2021training, uesato2022solving, saha2023branch, havrilla2024teaching}. Both \cite{cobbe2021training} and \cite{uesato2022solving} leveraged an external classifier to rank LLM outputs. Specifically, in both setups the authors proposed to generate many candidate solutions and select the one ranked highest by the verifier. The authors demonstrated the applicability of  using such verifiers in solving math word problems~\citep{gsm8k}. \cite{qin2023large} demonstrated that LLMs can serve as efficient text rankers when considering pairwise ranking. 

Another line of work leveraged LLMs to evaluate the quality of smaller models~\citep{saha2023branch, dubois2024alpacafarm, zheng2024judging,Oren:2024}. Although providing a promising alternative, such evaluation suffers from biases in the larger model~\citep{zheng2024judging} and reliance on hand-designed evaluation plans that impact the method’s ability to generalize~\citep{liu2023gpteval}. Large models also serve as verifiers of small ones in a speculative decoding setup, with the goal of speeding-up LLM generation~\citep{leviathan2023fast,sehoon2023speculative,chen2023accelerating}. 
It is also common to distill knowledge from a large model into a smaller one in order to improve efficiency \citep{hinton2015distilling,sanh2019distilbert,xu2024survey}, see \citet{treviso-etal-2023-efficient} for a survey on efficient methods in NLP.

In this work, we explore the potential of LLMs as selectors of the best output of a smaller model in a fixed budget setup. Similarly to ours, \cite{li2024common} found that smaller sized LMs ($7$B parameters) already exhibit strong mathematical abilities when selecting the best response from $k$ different generations. When considering code generation models,~\cite{alphacode2} presented impressive results on challenging coding contests tasks while generating $1$M samples, and later on filtering and ranking them using Gemini-Pro LLM~\citep{team2023gemini}. \cite{dou2024stepcoder} proposed a method to improve code-generation models by learning a policy model using reinforcement learning methods. Lastly, \citet{shi2022natural} and \citet{ni2023lever} used execution feedback in order to filter code-generations, while \citet{shi2022natural} used non-learned approaches, \citet{ni2023lever} trained an external verifier on top of the generation and the execution feedback.

\section{Discussion \& Limitations}

Our results show that using smaller models with the same amount of compute can improve LLM code-generation performance. An interesting question we do not fully address is whether, given enough compute, the larger models will overtake the smaller ones, or perhaps they will all saturate at a similar performance level at some point. Our HumanEval and MBPP results seem to slightly support the latter hypothesis (as all models begin to saturate, see \cref{fig:human_eval,fig:mbpp}). However, unfortunately, due to compute constraints, our setting is restricted to exploring only a limited number of generations per model.\footnote{\label{fn:prices}For instance, generating $1,000$ answers for the $5,000$ examples of the APPS benchmark with a $7$B model takes about $20$ days using a node of $8$ A$100$ GPUs.} We note that despite this limitation, in practice, due to these costs our conclusions apply to most practical use-cases.
We defer more expensive experiments to future work.

\section{Conclusion}
\label{sec:con}
In this work, we compared large language models with smaller-sized models under fixed budget constraints (i.e., FLOPs and wall-time). We evaluated the models using execution-based code-generation tasks, which provide access to unit-tests. Our findings reveal that generating multiple outputs from a $13$B model may lead to gains of up to 15\% over a single generation from a  $70$B model across five tasks. This highlights the potential of using smaller models instead of larger ones. In scenarios where unit tests or other solution verifiers are unavailable, we explored a simple ranking-based approach for candidate selection. We found the proposed ranking approach falls short in performance compared to a single output from the larger model. Our findings emphasize the importance of studying approaches for ranking LLM outputs, which hold great potential to not only improve model performance but also improve budget allocation. To further enhance this research direction we release over $1$M samples from the Code Llama 7B models spanning both HumanEval and MBPP benchmarks.

\section{Acknowledgments}
We thank Miri Varshavsky Hassid for the great feedback and moral support.

\bibliography{colm2024_conference}
\bibliographystyle{colm2024_conference}
\newpage

\appendix 
\section{Llama-3 Results}
\label{app:llama_3}
\begin{figure}[t]
     \centering
     \begin{subfigure}[b]{0.32\textwidth}
         \centering
         \includegraphics[trim={0.7cm 0.3cm 0.7cm 0.7cm},clip,width=\textwidth]{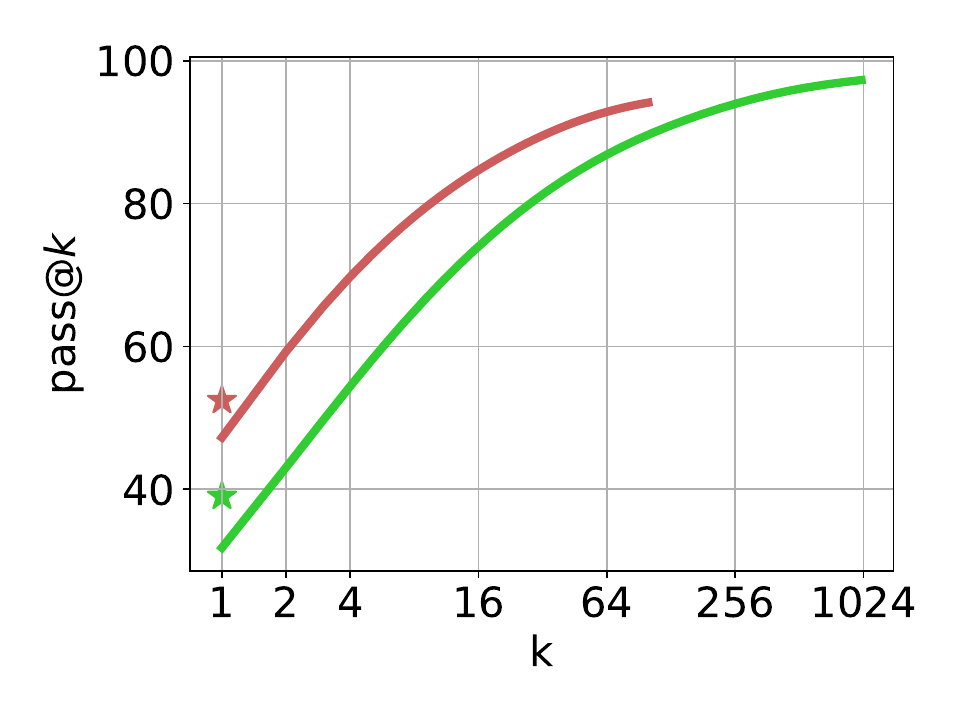} 
         \caption{\label{fig:llama_3_human_eval_k}}
     \end{subfigure}
     \hfill
     \begin{subfigure}[b]{0.32\textwidth}
         \centering
         \includegraphics[trim={0.7cm 0.3cm 0.7cm 0.7cm},clip,width=\textwidth]{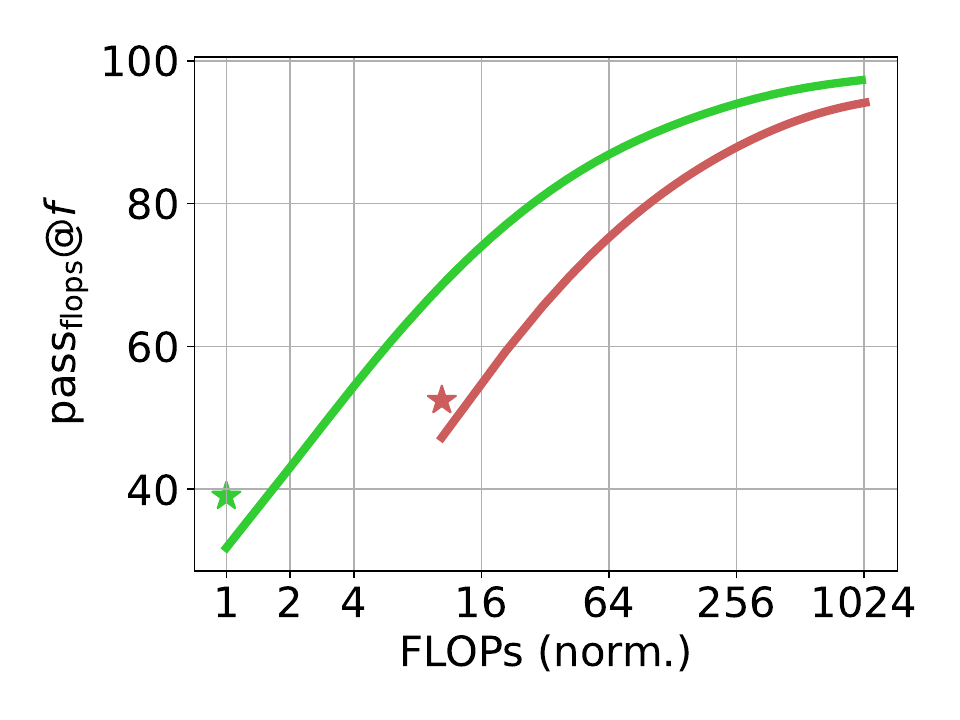} 
         \caption{\label{fig:llama_3_human_eval_flops}}
     \end{subfigure}
     \hfill
     \begin{subfigure}[b]{0.32\textwidth}
         \centering
         \includegraphics[trim={0.7cm 0.3cm 0.7cm 0.7cm},clip,width=\textwidth]{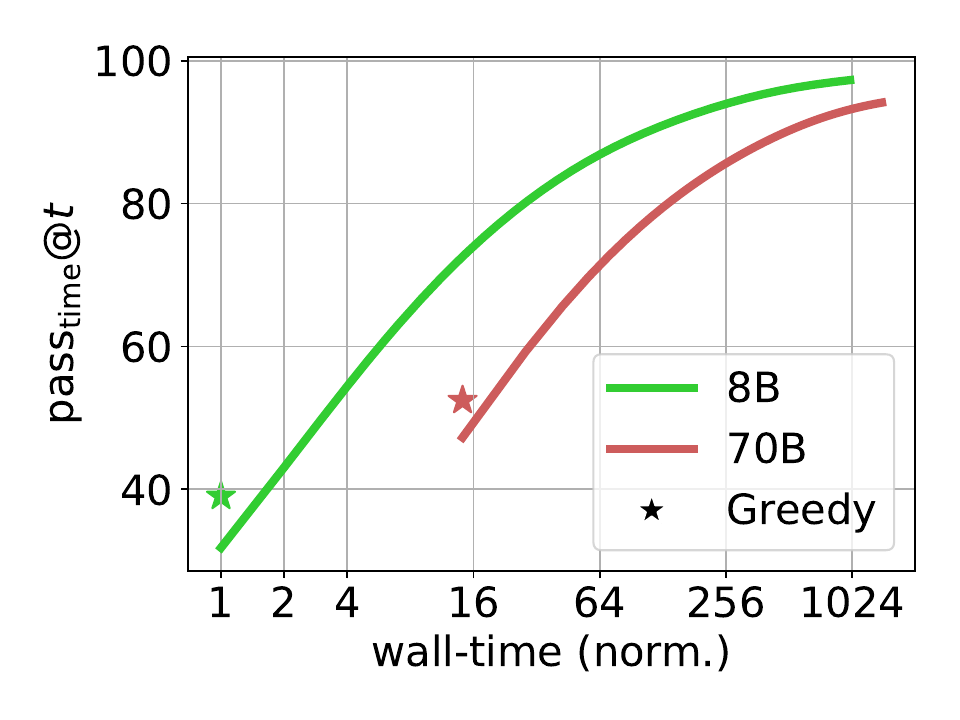} 
         \caption{\label{fig:llama_3_human_eval_time}}
     \end{subfigure}
     \caption{Llama-3 performance vs. compute for the HumanEval benchmark. The $70$B model performs better in general (\cref{fig:llama_3_human_eval_k}), but under a fixed compute budget (\cref{fig:llama_3_human_eval_flops,fig:llama_3_human_eval_time}), the $8$B model substantially outperforms the larger one.
\label{fig:llama_3_human_eval}}
\end{figure}

\begin{figure}[t]
     \centering
     \begin{subfigure}[b]{0.32\textwidth}
         \centering
         \includegraphics[trim={0.7cm 0.3cm 0.7cm 0.7cm},clip,width=\textwidth]{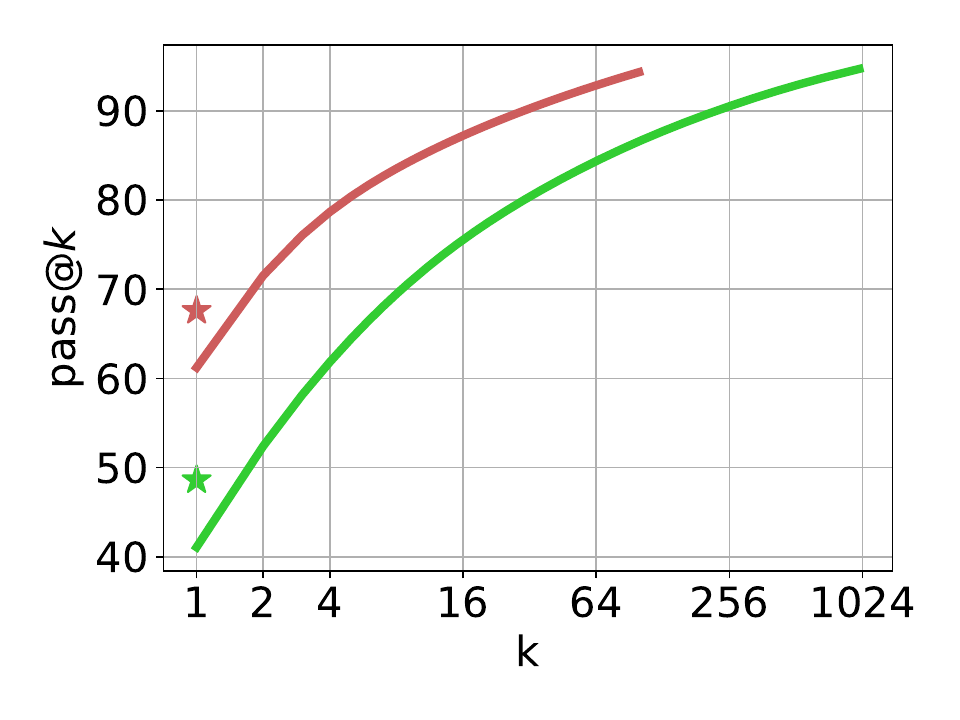}
         \caption{\label{fig:llama_3_mbpp_k}}
     \end{subfigure}
     \hfill
     \begin{subfigure}[b]{0.32\textwidth}
         \centering
         \includegraphics[trim={0.7cm 0.3cm 0.7cm 0.7cm},clip,width=\textwidth]{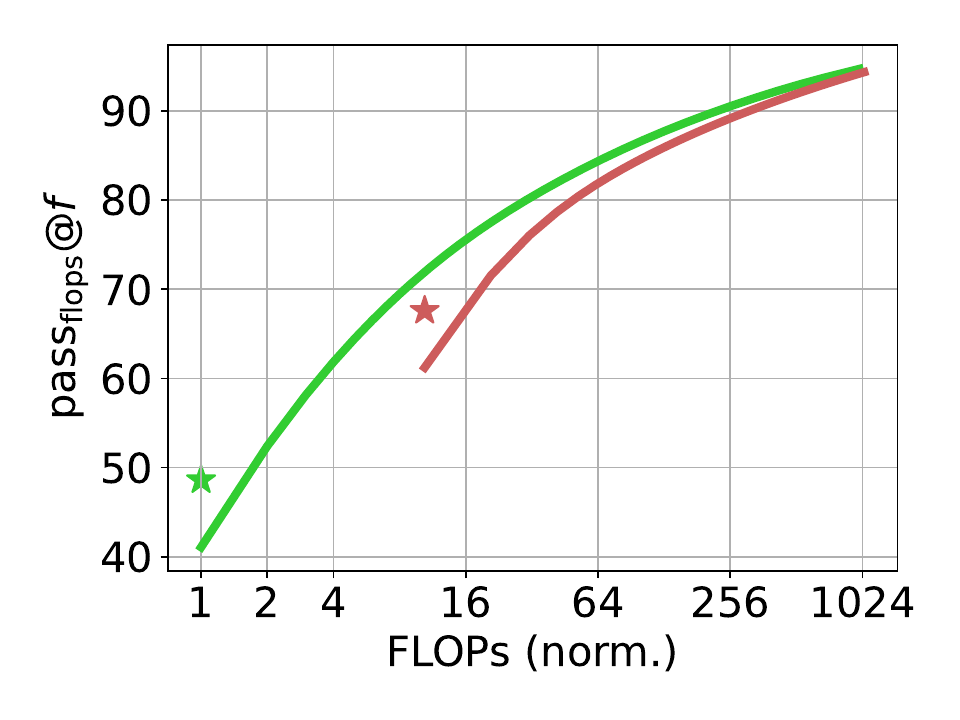}
         \caption{\label{fig:llama_3_mbpp_flops}}
     \end{subfigure}
     \hfill
     \begin{subfigure}[b]{0.32\textwidth}
         \centering
         \includegraphics[trim={0.7cm 0.3cm 0.7cm 0.7cm},clip,width=\textwidth]{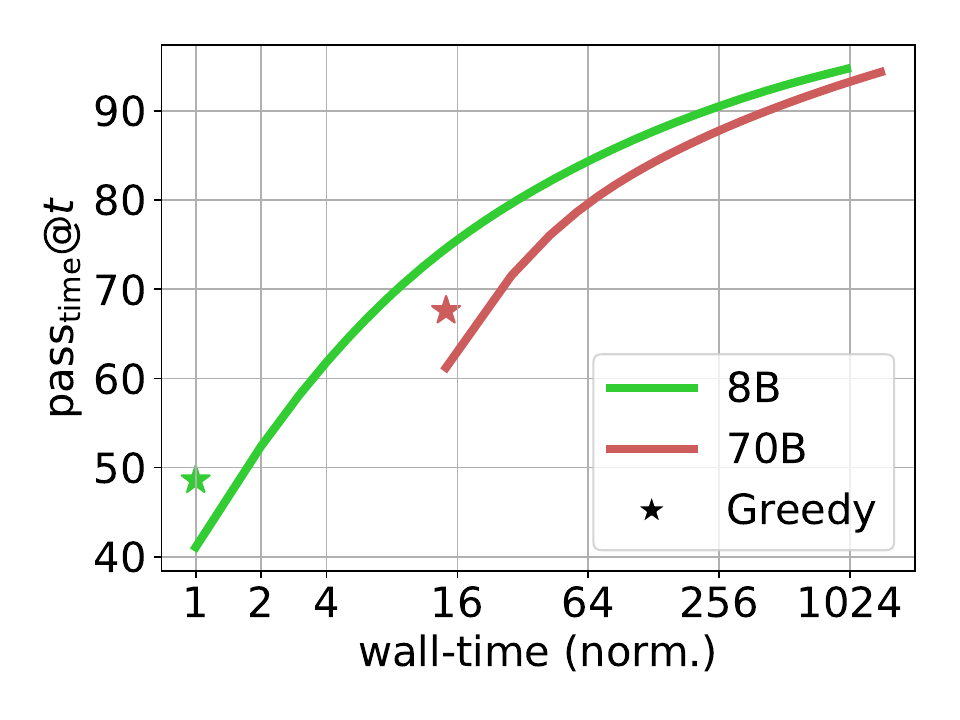}
         \caption{\label{fig:llama_3_mbpp_time}}
     \end{subfigure}
     \caption{Llama-3 performance vs.~compute for the MBPP benchmark. As in HumanEval~(\cref{fig:llama_3_human_eval}), larger models perform better as a function of $k$ (\cref{fig:llama_3_mbpp_k}), but worse under a fixed compute budget (\cref{fig:llama_3_mbpp_flops,fig:llama_3_mbpp_time}). %
\label{fig:llama_3_mbpp}}
\end{figure}

We present Llama-3 results for the HumanEval and MBPP benchmarks in \cref{fig:llama_3_human_eval,fig:llama_3_mbpp}, respectively.

\section{Detailed \passk Results}
\label{app:full_results}

In \cref{tab:full_human_eval,tab:full_mbpp,tab:full_apps} presents precise \passk results for the datasets examined (HumanEval, MBPP and APPS, respectively). Due to the infeasibility of reporting results for all $k$, we provide results for selected $k$ values. Nevertheless, it is important to note that all relevant $k$ values were calculated and used in the computation of the figures.

\begin{table}
\small
\caption{\label{tab:full_human_eval} Precise models' \passk results for several $k$ values over the HumanEval benchmark.}
\begin{center}
\begin{tabular}{l|cccccccccc}
\toprule

\diagbox[innerwidth=\textwidth*1/4]{Model}{$k$ for \passk} & 1 & 2 & 4 & 16 & 64 & 128 & 256 & 500 & 1000 \\
\midrule
Code Llama $7$B  & 28.2 & 38.5 & 48.9 & 68.7 & 83.9 & 89.0 & 92.7 & 95.0 & 96.3 \\
Code Llama $13$B & 32.7 & 44.4 & 56.2 & 77.0 & 89.0 & 92.3 & 94.8 & 96.5 & --.- \\
Code Llama $34$B & 38.6 & 51.4 & 63.4 & 81.8 & 91.7 & 94.3 & --.- & --.- & --.- \\
Code Llama $70$B & 46.7 & 61.1 & 73.2 & 87.3 & 94.4 & --.- & --.- & --.- & --.- \\
\midrule
Llama-3 $8$B & 31.8 & 43.1 & 54.4 & 74.0 & 86.9 & 90.9 & 94.0 & 96.0 & 97.3 \\
Llama-3 $70$B & 47.3 & 59.3 & 69.8 & 84.7 & 92.8 & --.- & --.- & --.- & --.- \\
\bottomrule
\end{tabular}
\end{center}
\end{table}

\begin{table}
\small
\caption{\label{tab:full_mbpp} Precise models' \passk results for several $k$ values over the MBPP benchmark.}
\begin{center}
\begin{tabular}{l|cccccccccc}
\toprule

\diagbox[innerwidth=\textwidth*1/4]{Model}{$k$ for \passk} & 1 & 2 & 4 & 16 & 64 & 128 & 256 & 500 & 1000 \\
\midrule
Code Llama $7$B  & 37.3 & 48.0 & 57.3 & 71.6 & 81.1 & 84.8 & 87.9 & 90.3 & 92.4 \\
Code Llama $13$B & 42.3 & 53.4 & 62.5 & 75.3 & 84.0 & 87.4 & 90.4 & 92.8 & --.- \\
Code Llama $34$B & 49.2 & 60.1 & 68.6 & 79.2 & 85.8 & 88.7 & --.- & --.- & --.- \\
Code Llama $70$B & 57.3 & 67.7 & 74.8 & 83.7 & 90.0 & --.- & --.- & --.- & --.- \\
\midrule
Llama-3 $8$B & 41.1 & 52.4 & 61.8 & 75.5 & 84.4 & 87.7 & 90.5 & 92.8 & 94.7 \\
Llama-3 $70$B & 61.2 & 71.5 & 78.7 & 87.2 & 92.9 & --.- & --.- & --.- & --.- \\
\bottomrule
\end{tabular}
\end{center}
\end{table}

\begin{table}
\small
\caption{\label{tab:full_apps} Precise models' \passk results for several $k$ values over the different splits of the APPS benchmark.}
\begin{center}
\begin{tabular}{l|cccccccccc}
\toprule
\diagbox[innerwidth=\textwidth*1/4]{Model}{$k$ for \passk} & 1 & 2 & 4 & 16 & 32 & 64 & 128 & 256 & 500 \\
\midrule
\multicolumn{10}{c}{APPS-introductory} \\
\midrule
Code Llama $7$B  & 4.5 & 7.7 & 12.1 & 23.0 & 28.6 & 34.0 & 39.3 & 44.4 & 49.0 \\
Code Llama $13$B & 10.3 & 16.3 & 23.1 & 36.3 & 42.3 & 47.8 & 52.8 & --.- & --.- \\
Code Llama $34$B & 12.6 & 19.7 & 27.4 & 40.8 & 46.8 & 52.5 & --.- & --.- & --.- \\
Code Llama $70$B & 22.6 & 30.5 & 38.1 & 52.5 & 58.5 & --.- & --.- & --.- & --.- \\
\midrule
\multicolumn{10}{c}{APPS-interview} \\
\midrule
Code Llama $7$B  & 0.9 & 1.6 & 2.7 & 6.1 & 8.3 & 10.8 & 13.6 & 16.4 & 19.2 \\
Code Llama $13$B & 2.2 & 3.8 & 5.9 & 11.6 & 14.8 & 18.3 & 21.9 & --.- & --.- \\
Code Llama $34$B & 2.9 & 4.9 & 7.6 & 14.4 & 18.1 & 22.0 & --.- & --.- & --.- \\
Code Llama $70$B & 5.5 & 8.7 & 12.6 & 21.8 & 26.6 & --.- & --.- & --.- & --.- \\
\midrule
\multicolumn{10}{c}{APPS-competition} \\
\midrule
Code Llama $7$B  & 0.1 & 0.3 & 0.5 & 1.6 & 2.6 & 4.1 & 6.1 & 8.6 & 11.6 \\
Code Llama $13$B & 0.6 & 1.2 & 2.1 & 5.3 & 7.8 & 10.9 & 14.4 & --.- & --.- \\
Code Llama $34$B & 0.7 & 1.3 & 2.3 & 6.1 & 9.0 & 12.6 & --.- & --.- & --.- \\
Code Llama $70$B & 1.7 & 3.0 & 4.9 & 11.1 & 15.3 & --.- & --.- & --.- & --.- \\
\bottomrule
\end{tabular}
\end{center}
\end{table}

\end{document}